\documentclass[final,10pt]{iopart}

\usepackage{amssymb}
\usepackage[]{hyperref}

\usepackage{graphicx}
\usepackage{mathrsfs}

\newcommand{\D}{\mathrm{D}}
\newcommand{\eqref}[1]{\eref{#1}}

\usepackage{xcolor}
\hypersetup{
    colorlinks,
    linkcolor={blue},
    citecolor={blue},
    urlcolor={blue}
}

\makeatletter 

\newcommand{\insertheaderrule}{\rlap{\rule[-.3\normalbaselineskip]{\textwidth}{.4pt}}}
\def\@oddhead{\insertheaderrule{\footnotesize\sffamily\leftmark}\hfil{\footnotesize\sffamily\thepage}}%

\renewcommand{\title}{\@ifnextchar[{\@stitle}{\@ftitle}}
\def\@stitle[#1]#2{\markboth{#1}{#1}%
    \thispagestyle{myheadings}%
    \vspace*{3pc}{\exhyphenpenalty=10000\hyphenpenalty=10000 
    \huge\raggedright\noindent
    \bf\textsf#2\par}}
\def\@ftitle#1{\markboth{#1}{#1}%
    \thispagestyle{myheadings}%
    \vspace*{3pc}{\exhyphenpenalty=10000\hyphenpenalty=10000 
    \huge\raggedright\noindent
    \bf\textsf{#1}\par}}

\renewenvironment{abstract}{%
      \vspace{16pt plus3pt minus3pt}
      \begin{indented}
      \item[]{\normalsize\bfseries \textsf{\abstractname}\\}\quad\rm\ignorespaces\normalsize} 
      {\end{indented}\if@titlepage\newpage\else\vspace{36pt}\fi}

\renewcommand\section{\@startsection {section}{1}{\z@}%
                   {-3.5ex \@plus -1ex \@minus -.2ex}%
                   {2.3ex \@plus.2ex}%
                   {\reset@font\normalsize\bfseries\sffamily\raggedright}}
\renewcommand\subsection{\@startsection{subsection}{2}{\z@}%
                   {-3.25ex\@plus -1ex \@minus -.2ex}%
                   {1.5ex \@plus .2ex}%
                   {\reset@font\normalsize\itshape\sffamily\raggedright}}
\renewcommand\subsubsection{\@startsection{subsubsection}{3}{\z@}%
                                     {-3.25ex\@plus -1ex \@minus -.2ex}%
                                     {-1em \@plus .2em}%
                                     {\reset@font\normalsize\itshape\sffamily}}
\renewcommand\paragraph{\@startsection{paragraph}{4}{\z@}%
                                    {3.25ex \@plus1ex \@minus.2ex}%
                                    {-1em}%
                                    {\reset@font\normalsize\itshape\sffamily}}
      
\makeatother 
\def\enquote#1{`#1'}
      
\begin{document}

\title{Slicing conditions for axisymmetric gravitational collapse of Brill waves}
\author{\large\textsf{Anton Khirnov and Tom\'{a}\v{s} Ledvinka}}
\address{Institute of Theoretical Physics, Faculty of Mathematics and Physics, Charles University, 
V~Hole\v{s}ovi\v{c}k\'{a}ch~2, Prague, 18000, Czech Republic}
{\vspace*{15pt}\address{E-mail: \mailto{\mailto{anton@khirnov.net}, \mailto{Tomas.Ledvinka@mff.cuni.cz}}}}

\begin{abstract}
  In numerical relativity, spacetimes involving compact strongly gravitating objects are constructed as numerical
  solutions of Einstein's equations. Success of such a process strongly depends on the availability of appropriate
  coordinates, which are typically constructed dynamically. A very robust coordinate choice is a so-called moving
  puncture gauge, commonly used for numerical simulations of black hole spacetimes. Nevertheless it is known to fail
  for evolving near-critical Brill wave data. We construct a new `quasi-maximal' slicing condition and
  demonstrate that it exhibits better behavior for such data. This condition is based on the 1+log slicing with an
  additional source term derived from maximal slicing. It is relatively simple to implement in existing moving puncture
  codes and computationally inexpensive. We also illustrate the properties of constructed spacetimes based on
  gauge-independent quantities in compactified spacetime diagrams. These invariants are also used to show how created
  black holes settle down to a Schwarzschild black hole.
\end{abstract}


\section{Introduction}
During the last decade numerical relativity became tremendously successful at simulating many astrophysically important
processes in which general relativity plays a crucial role. A significant portion of this progress can be attributed to
the discovery of coordinates suitable for such wildly dynamic spacetime geometries. Present codes can handle mergers
both of purely vacuum black hole binaries and neutron star binaries (followed by a collapse). It may thus seem
surprising that a certain theoretically important situation \textemdash{} gravitational wave packet collapse
\textemdash{} still defies these codes.

Numerical treatment of gravitational waves collapsing to form a black hole dates back to the early age of numerical
relativity. Eppley in his pioneering work \cite{eppley_1977} successfully constructed what is known as the Brill waves
\cite{Brill_1959} \textemdash{} a family of vacuum initial data describing axisymmetric wave packets at the moment of
time symmetry \textemdash{} and located apparent horizons in them. Another breakthrough was the paper of Abrahams and
Evans \cite{Abrahams_1993} which used another family of axisymmetric data called the Teukolsky waves
\cite{Teukolsky_1982}.  They found evidence of critical behavior \textemdash{} discrete self-similarity and power-law
scaling of various quantities close to the threshold of black hole formation \textemdash{} analogous to the results of
Choptuik for a scalar field \cite{Choptuik_1993,gundlach_2007}.

Since then, as computer performance and numerical methods improved, several attempts were made to investigate the
nonlinear regime of the Brill data. Alcubierre et al.\ \cite{alcubierre_2000} used a combination of the BSSN evolution
system and maximal slicing to put rough bounds on the critical value of the amplitude parameter. That result was
confirmed in \cite{garfinkle_2001} using a mixed elliptic-hyperbolic reduction of the evolution equations, again with
maximal slicing. Ultimately, insufficient performance of contemporary computers constrained what could be achieved with
these attempts.

As the moving puncture gauge \cite{bona_1995,alcubierre_2003} showed its strength for black hole mergers, Hilditch et
al.\ \cite{Hilditch_2013} attempted to use it also for gravitational wave collapse. One of their discoveries was the
pathological behavior of this gauge for near-critical Brill waves, producing what they conjectured to be coordinate
singularities.  Recently \cite{Hilditch_2016,hilditch_2017} they managed to avoid these problems with a new
pseudospectral code using a generalized harmonic formulation.

In this paper, we follow up on the work of \cite{Hilditch_2013}, using the same initial data and similar numerical
techniques (although a completely different code). We analyze the way in which the moving puncture gauge \textemdash{} a
combination of the \enquote{1+log} slicing and the \enquote{$\Gamma$-driver} shift condition \textemdash{} breaks, and
modify the slicing condition. Our modification takes the form of an extra source term pushing the constant-time
hypersurfaces closer towards maximal slicing, which is known to guarantee smooth solutions and has shown promising
results in earlier work \cite{alcubierre_2000,garfinkle_2001,rinne_2008}. The result is a new slicing condition
\textemdash{} which we call \enquote{quasi-maximal slicing} \textemdash{} that no longer exhibits the problems reported
in \cite{Hilditch_2013}.  While quasi-maximal slicing involves solving an elliptic equation, solver accuracy only
affects the distance from the preferred gauge and does not give rise to ADM constraint violations \textemdash{} in
contrast to the usual implementations of maximal slicing.

Using this slicing, we illustrate features of gravitational wave collapse that are present even farther away from the
critical amplitude, such as formation of trapped surfaces and the event horizon, and construct compactified diagrams of
considered spacetimes. These were previously not available, since earlier techniques are not able to handle long-term
simulations easily.

This paper is laid out as follows. In Section \ref{sec:continuum} we briefly summarize the continuum equations on
which our numerics is based. In Section \ref{sec:qms} we introduce our modified slicing condition. Section
\ref{sec:numerics} describes the key points of the numerical codes used and why we trust their output. Finally in
Section \ref{sec:results} we present the results for long-term evolution of the considered class of initial data. We use
geometrized units $G=c=1$ in this paper.
\section{Field equations}
\label{sec:continuum}
\subsection{Brill wave initial data}
\label{sec:initial_data}

The Brill waves \cite{Brill_1959} are a family of vacuum axially symmetric initial data defined by the spatial metric
in cylindrical coordinates $\left\{ \rho, z, \varphi \right\}$
\begin{equation}
  \gamma_{ij} = \psi^4 \left[ e^{2q}\left( \mathrm{d}\rho^2 + \mathrm{d}z^2 \right) + \rho^2
    \mathrm{d}\varphi^2 \right],
  \label{eq:brillwaves}
\end{equation}
and the condition of time symmetry, which implies initially vanishing extrinsic curvature. Here $\psi$ is the
conformal factor and $q=q(\rho, z)$ is the so-called \enquote{seed function}, which needs to have certain regularity and
decay properties \cite{Brill_1959}, but can otherwise be chosen arbitrarily. For ease of comparison, we use the same
form of $q$ as was used in \cite{Hilditch_2013}, specifically
\begin{equation}
  q(\rho, z) = A \left(\frac{\rho}{\sigma}\right)^2 e^{-(\rho^2 + z^2) / \sigma^2}.
  \label{eq:brill_q}
\end{equation}
The parameter $A$ determines the amplitude of the waves, with $A = 0$ being flat space. We only consider non-negative
values of $A$ in this paper, though it is worth noting that $A<0$ also produces valid initial data. The critical point
$A^*$ is the smallest value of $A$ for which a gravitational singularity is formed. For the above choice of $q$ it is
bracketed by our simulations to $A^* \in  \left[ 4.69, 4.7 \right]$, which is compatible with tighter bounds given in
\cite{hilditch_2017}.

The scale $\sigma$ (with dimension of length) fixes the units of all physical quantities. In our code we set $\sigma =
1$, so all numerical values of time, length or mass are assumed to be in units of $\sigma$. This allows us to
compare our results directly with \cite{Hilditch_2013}, where the same choice is made. For quantities that do not
clearly have dimension of length we print the appropriate power of $\sigma$ explicitly.

Due to time symmetry the momentum constraints are trivial, so to construct the data we only need to solve the
Hamiltonian constraint. It reduces to
\begin{equation}
  \Delta \psi + \frac{1}{4} \left( \partial_{\rho\rho}q + \partial_{zz}q \right) \psi = 0,
  \label{eq:brill_ham}
\end{equation}
a linear elliptic equation for the conformal factor $\psi$, where $\Delta$ is the flat space Laplacian. Our code for
solving this equation is described in Section \ref{ssec:implementation}.
\subsection{Evolution system}
\label{sec:evol_system}
We follow the standard 3+1 splitting procedure, with the 4-dimensional metric $g_{\mu\nu}$ decomposed into the
spatial metric $\gamma_{ij}$, the lapse $\alpha$ and the shift vector $\beta^i$
\begin{equation}
  \mathrm{d}s^2 = \left( -\alpha^2 + \beta_i \beta^i \right)\mathrm{d}t^2 + 2\beta_i \mathrm{d}t \mathrm{d}x^i +
  \gamma_{ij}\mathrm{d}x^i \mathrm{d}x^j.
  \label{eq:metric_decomposition}
\end{equation}
The vacuum Einstein equations can then be written as a set of evolution equations
\numparts
  \begin{eqnarray}
    \left( \partial_t - \mathcal{L}_\beta \right)\gamma_{ij} &= -2\alpha K_{ij},
    \label{eq:gamma_dot}\\
    \left( \partial_t - \mathcal{L}_\beta \right)K_{ij} &= -\D_i \D_j \alpha + \alpha \left[ R_{ij} + K K_{ij} - 2 K_{ik} K_{j}^k \right],
      \label{eq:k_dot}
    \end{eqnarray}
\endnumparts
and a set of constraints
\numparts
  \begin{eqnarray}
    {}^{(3)}R + K^2 - K_{ij} K^{ij} &= 0,
    \label{eq:ham_constraint}\\
    \D_j K^{ij} - \D^i K &= 0.
    \label{eq:mom_constraint}
  \end{eqnarray}
\endnumparts
Here $K_{ij}$, defined by \eqref{eq:gamma_dot}, is the extrinsic curvature of the spatial slices, $\mathcal{L}_\beta$ is
the Lie derivative along $\beta^i$, $\D_i$, $R_{ij}$ and ${}^{(3)}R$ are respectively the covariant derivative, the Ricci tensor
and the scalar curvature associated with $\gamma_{ij}$.

Introducing new evolved variables
\numparts
  \begin{eqnarray}
    \varphi &= \mathrm{det}\left( \gamma_{ij} \right)^{-\frac{1}{6}},
    \label{eq:phi_def}\\
    K &= \gamma_{ij}K^{ij},
    \label{eq:trk_def}\\
    \bar{\gamma}_{ij} &= \varphi^2 \gamma_{ij},
    \label{eq:gammabar_def}\\
    \bar{A}_{ij} &= \varphi^2 \left( K_{ij} - \frac{1}{3} K\gamma_{ij} \right),
    \label{eq:abar_def}\\
    \bar{\Gamma}^i &= \bar{\gamma}^{jk}\bar{\Gamma}_{jk}^i,
    \label{eq:xbar_def}
  \end{eqnarray}
\endnumparts
where $\bar{\Gamma}_{jk}^i$ are the Christoffel symbols associated with $\bar{\gamma}_{ij}$, and using the constraints,
we can write the evolution equations as
\numparts
  \begin{eqnarray}
    \left( \partial_t -\mathcal{L}_\beta \right)\varphi &= \frac{1}{3}\varphi\alpha K,
    \label{eq:phi_dot}\\
    \left( \partial_t -\mathcal{L}_\beta \right)\bar{\gamma}_{ij} &= -2\alpha \bar{A}_{ij},
    \label{eq:gammabar_dot}\\
    \left( \partial_t -\mathcal{L}_\beta \right)K &= -\D^i \D_i \alpha + \alpha \left( A_{ij}A^{ij} + \frac{1}{3}K^2 \right),
    \label{eq:trk_dot}\\
    \left( \partial_t -\mathcal{L}_\beta \right)\bar{A}_{ij} &= \varphi^2 \left[ -\D_i \D_j \alpha + \alpha R_{ij} \right]^{\mathrm{TF}}
      + \alpha\left( K \bar{A}_{ij} - 2 \bar{A}_{ik} \bar{A}_{j}^k \right),
    \label{eq:abar_dot}\\
    \left( \partial_t -\mathcal{L}_\beta \right)\bar{\Gamma}^i &=
          \bar{\gamma}^{jk}\partial_j \partial_k \beta^i + \frac{1}{3} \bar{\gamma}^{ij}\partial_j \partial_k \beta^k - 2\bar{A}^{ij}\partial_j \alpha\\
          &\nonumber\quad+2\alpha\left( \bar{\Gamma}_{jk}^i \bar{A}^{jk} + 6\bar{A}^{ij} \partial_j \varphi - \frac{2}{3} \bar{\gamma}^{ij}\partial_j K\right),
    \label{eq:xbar_dot}
  \end{eqnarray}
\endnumparts
where TF denotes the trace-free part of the bracketed expression. The conformally related metric $\bar{\gamma}_{ij}$ and
its inverse $\bar{\gamma}^{ij}$ are used to raise and lower indices on all the quantities with a bar. Terms involving
$\D_i$, $R_{ij}$ and ${}^{(3)}R$ are now assumed to be computed from $\bar{\gamma}_{ij}$ and $\varphi$ and their
derivatives in a straightforward manner.

This is known as the BSSN \cite{Shibata_1995,Baumgarte_1998} formulation, which (given appropriate gauge choice) is
known to be strongly hyperbolic, and has a solid track record in black hole simulations.

One of the slicing conditions we consider is the maximal slicing, defined by requiring the volume elements associated
with normal observers to remain constant. This implies that the trace of the extrinsic curvature $K$ is identically zero
at all times. Provided that the initial slice satisfies $K = 0$, equation \eqref{eq:trk_dot} immediately gives us
\begin{equation}
  \partial_t K = -\D^i \D_i \alpha + K_{ij}K^{ij}\alpha = 0,
  \label{eq:maximal_slicing}
\end{equation}
a linear elliptic equation for the lapse $\alpha$ that needs to be solved at each step of the evolution. For its many
desirable properties \cite{alcubierre_2008}, this slicing has been used since early times of numerical relativity, and
specifically for Brill waves e.g. in \cite{alcubierre_2000,garfinkle_2001}. Its main disadvantage is the fact that
solving an elliptic equation at each time step is usually impractical in 3D and tends to be very resource-intensive even
in 2D.

According to \cite{Gundlach_2006}, maximal slicing is only well-posed when the constraint $K = 0$ is enforced, which is
easily achieved with BSSN by just not evolving $K$.

The other common lapse choice we make use of is the 1+log slicing \cite{bona_1995}, given by a hyperbolic evolution
equation for the lapse
\begin{equation}
  \left( \partial_t - \mathcal{L}_\beta \right)\alpha = -2\alpha K.
  \label{eq:1+log}\\
\end{equation}
This condition is intended to approximate the main advantages of the maximal slicing at a lower computational cost.

For the shift we mainly use the $\Gamma$-driver condition (original version introduced in \cite{alcubierre_2003}, the
flavor we use is from \cite{van_meter_2006})
\begin{equation}
  \left( \partial_t -\mathcal{L}_\beta \right)\beta^i = \xi \bar{\Gamma}^i - \eta \beta^i,
\end{equation}
a hyperbolic equation related to elliptic minimal-distortion conditions. Here the dimensionless parameter $\xi$
determines the velocities of the longitudinal and transversal gauge waves, while $\eta$ is a damping parameter (with a
dimension of inverse length) that seems to be necessary to avoid shock formation in the shift. In some cases we also set
the shift to be identically zero.

Together 1+log slicing and $\Gamma$-driver shift are known as the moving puncture gauge and are widely used in numerical
relativity, among other thing for black-hole spacetimes.
\subsection{Invariants}
\label{sec:invariants}
To extract coordinate-independent information from our simulations we look at spacetime invariants. Since the end-state
of the super-critical runs has to be a Schwarzschild black hole, the invariants' behavior in Schwarzschild is of special
interest to us.

The most prominent scalars are those constructed from curvature, e.g. the Kretschmann scalar $\mathcal{K} =
R_{\mu\nu\alpha\beta} R^{\mu\nu\alpha\beta}$ (where $R_{\mu\nu\alpha\beta}$ is the four-dimensional Riemann tensor). For
a Schwarzschild black hole with mass $M$ we have
\begin{equation}
  \mathcal{K}_{\mathrm{Schwarzschild}} = \frac{48 M^2}{R^6},
  \label{eq:kretschmann_schwarzschild}
\end{equation}
where $R$ is the usual areal radius. This allows us to define the \enquote{Kretschmann mass} as
\begin{equation}
  M_{\mathcal{K}}(R) = \sqrt{\frac{\mathcal{K} R^6}{48}},
  \label{eq:mass_kretschmann}
\end{equation}
where $\mathcal{K}$ and $R$ are evaluated at some point in the equatorial plane $z=0$.

One practical issue with $\mathcal{K}$ is that it contains second derivatives of the evolved quantities and its
stationary limit (on which the mass estimate \eqref{eq:mass_kretschmann} is based) has a rather fast $1 / R^6$ falloff.
When the evolved metric contains noise \textemdash{} e.g. caused by reflections on the grid refinement boundaries
\textemdash{} the spurious terms decay less steeply and can become of comparable size to the stationary value.

Other invariants, which are simpler and easier to calculate, can be obtained from the fact that our spacetimes are
axially symmetric. That implies the existence of an angular Killing vector $\eta^\mu =
\frac{\partial}{\partial\varphi}^\mu$.  When appropriately scaled, its norm is the circumferential radius
\begin{equation}
  \bar{\rho}^2 =\eta_\mu \eta^\mu = \gamma_{yy} x^2,
  \label{eq:circ_radius}
\end{equation}
where the second equality holds in the $y=0$ plane. Clearly in spherical symmetry we have $\bar{\rho}^2 = R^2$ in the
equatorial plane. Another related scalar we can construct is the norm of the gradient of $\bar{\rho}^2$
\begin{equation}
  4\bar{\rho}'^2 = \left|\nabla \bar{\rho}^2\right|^2 = (\D_i \bar{\rho}^2) (\D^i\bar{\rho}^2) - 4 (K_{ij}\eta^i
  \eta^j)^2.
  \label{eq:rhot2}
\end{equation}
Note that despite the notation, $\bar{\rho}'^2$ can be negative \textemdash{} e.g. in Schwarzschild the
$R=\mathrm{const.}$ surfaces are spacelike above the horizon and timelike below it.

From the above two scalars we can construct a dimensionless quantity
\begin{equation}
  \chi = \frac{\bar{\rho}'^2}{\bar{\rho}^2}.
  \label{eq:chi}
\end{equation}
In our simulations we can use $\chi$ as a measure of spacetime deformation in the equatorial plane.
For the Schwarzschild solution $\chi = 1 - \frac{2M}{R}\sin ^2 \theta$, so in super-critical spacetimes the black hole
horizon will settle down to the $\chi=0$ hypersurface. Additionally, in stationary regions we can form another mass
estimate in the equatorial plane
\begin{equation}
  M_{\bar{\rho}} = \frac{\bar{\rho}^2 - \bar{\rho}'^2}{2\bar{\rho}}.
  \label{eq:mrho}
\end{equation}

In future studies of the critical collapse of axisymmetric gravitational waves, the quantity $\chi$ may be of interest,
since for a spacetime with discrete self-similarity such a dimensionless quantity should repeat with the same range of
values on each echoing period \cite{gundlach_2007}.
\section{Modified 1+log slicing}
\label{sec:qms}
Though maximal slicing appears to be well-behaved for near-critical Brill waves, implementing it in a numerical code
entails practical difficulties \textemdash{} solving elliptic equations at each time step is computationally demanding
and errors involved immediately lead to constraint violations. However, since the lapse is a gauge function, we do not
actually insist on it being exactly the maximal one. A \enquote{good-enough} approximation that still avoids the
problems of the 1+log slicing works just as well for our purposes.

In obtaining such an approximation, we start with the maximal slicing condition \eqref{eq:maximal_slicing} and take its
time derivative
\begin{equation}
  0 = \left( \partial_t - \mathcal{L}_\beta \right) \left[ \gamma^{ij}\D_i \D_j \alpha -
    K_{ij}K^{ij}\alpha \right],
  \label{eq:dt_maximal_slicing}
\end{equation}
which with some straightforward manipulations transforms into
\begin{eqnarray}
    &\gamma^{ij}\D_i \D_j \left[ (\partial_t - \mathcal{L}_\beta) \alpha \right] - K_{ij}K^{ij} \left[ (\partial_t -
      \mathcal{L}_\beta)\alpha \right] = \\
    \nonumber&-\left[ \left( \partial_t - \mathcal{L}_\beta \right)\gamma^{ij} \right]\D_i \D_j
      \alpha + \gamma^{ij}\left( \partial_t \Gamma_{ij}^k \right)\partial_k \alpha \\
    \nonumber&- \left( \gamma^{ij} \D_i \D_j \beta^k \right) \D_k \alpha - \beta^j R_{j}^i \D_i \alpha
    + \alpha \left( \partial_t - \mathcal{L}_\beta \right) \left(
      K_{ij}K^{ij} \right).
  \label{eq:qms1}
\end{eqnarray}
Now we replace the time derivative of the lapse with a new function $W$
\begin{equation}
  W = (\partial_t - \mathcal{L}_\beta)\alpha
  \label{eq:defw}
\end{equation}
and use the evolution equations to replace the time derivatives of $\gamma_{ij}$ and $K_{ij}$ with purely
spatial quantities known on one slice. We obtain
\begin{eqnarray}
  \label{eq:qms2}
    &\D^i \D_i W - K_{ij} K^{ij}W =\\
    \nonumber&-2\alpha K^{ij} \D_i \D_j \alpha + \gamma^{ij}\left( \partial_t
    \Gamma_{ij}^k \right)\partial_k \alpha - \left( \gamma^{ij} \D_i \D_j \beta^k \right) \D_k \alpha\\
    \nonumber&-\beta^j R_{j}^i \D_i \alpha +\alpha \left( 2 \dot{K_{ij}}K^{ij} + 4\alpha K_{j}^i K_{i}^{k} K_{k}^j \right),
\end{eqnarray}
where $\dot{K}_{ij}$ is a shorthand for the right-hand side of \eqref{eq:k_dot}. This is an elliptic equation for the
new function $W$, with the same structure as the maximal slicing condition \eqref{eq:maximal_slicing}, but a more
complicated right-hand side.

Our starting point \eqref{eq:dt_maximal_slicing} is nothing else than the demand that the acceleration of $K$ vanishes.
So if we now solved \eqref{eq:qms2} exactly at each time step and used its solution $W_e$ to evolve the lapse according to
\eqref{eq:defw}, then \textemdash{} assuming that initially $K=0$ and the lapse is maximal \textemdash{} we would get
precisely the maximal slicing (leaving aside the question of stability of such a system).  Of course in a numerical
simulation we will not have an exact solution and good approximations to it do not come cheaply. One can hope, however,
that even a rough approximation to the solution of \eqref{eq:qms2} can be exploited to get closer to maximal slicing.

The basic idea is then as follows \textemdash{} at each time step we compute $W_a$, a (very approximate) solution to
\eqref{eq:qms2}. Then we use this quantity as an extra source term in the 1+log slicing
\begin{equation}
  (\partial_t - \mathcal{L}_\beta)\alpha = -2\alpha K + \kappa W_a,
  \label{eq:qms}
\end{equation}
where $\kappa$ is a function of coordinate time used to switch between slicing conditions \textemdash{} we set it equal
to one when the extra term is to be active and smoothly send it to zero when we want to recover the original 1+log
slicing.

The motivation for this extra term is as follows. In the weak-field regime, the Bona-Mass\'o class of slicing conditions
yields a hyperbolic $(K, \alpha)$ subsystem \textemdash{} a scalar sector corresponding to so-called lapse gauge waves.
These exhibit the usual wave behavior and naturally drive weak-field configurations to $K=0$ as waves (assuming their
finite \enquote{energy}) disperse outwards. The aim of our modification \eqref{eq:qms} is to push the slicing towards
$K=0$ even in a non-linear regime.

It is known that for PDE systems such as BSSN we are not free to modify even the gauge conditions arbitrarily
\textemdash{} adding combinations of evolved functions and their derivatives may lead to loss of well-posedness of
the system. The source term added in \eqref{eq:qms} does not modify the principal part of the PDEs linearized
around flat space, so it does not affect the hyperbolicity of the PDE system. This follows from the fact that
\eqref{eq:qms2} has a trivial principal part \textemdash{} $\Delta W = 0$. Also \textemdash{} as described later
\textemdash{} in our approach we do not assume that the elliptic solver should provide a solution close to the continuum
limit.

Further in this paper we show numerical evidence that slicing condition \eqref{eq:qms} is not only stable for
small-amplitude Brill waves, but also that for larger amplitudes the extra term acts as a driver that pushes the slicing
closer to the maximal one, curing the problems of 1+log slicing. We call this condition the \enquote{quasi-maximal}
slicing.
\section{Numerics}
\label{sec:numerics}
\subsection{Implementation}
\label{ssec:implementation}

\paragraph*{Initial data}

Due to axial symmetry, the equation \eqref{eq:brill_ham} only needs to be solved in two dimensions. We use a
pseudo-spectral method, writing the conformal factor as a series \begin{equation}
  \psi\left( x^0, x^1 \right) = 1 + \sum_{k = 0}^{N-1} \sum_{l = 0}^{M-1} C_{kl} B_{k}^0 \left( x^0 \right) B_{l}^1 \left( x^1 \right).
  \label{eq:brill_spectral_expansion}
\end{equation}
Using \eqref{eq:brill_spectral_expansion} in \eqref{eq:brill_ham} and demanding that the equation be satisfied exactly at $NM$ collocation
points gives us $NM$ linear equations for the coefficients $C_{kl}$. We solve those using LU decomposition, which then allows us to
reconstruct $\psi$ at arbitrary points through \eqref{eq:brill_spectral_expansion}.

Specifically, we choose polar coordinates $\left\{ x^0, x^1 \right\} = \left\{ r, \theta \right\}$ and basis
functions \cite{Boyd_1987}
\numparts
  \begin{eqnarray}
    B_{k}^0 \left( r \right) &= \mathrm{SB}_{2k}\left( r \right) = \sin \left( \left( 2k + 1 \right)
    \mathrm{arccot}\frac{r}{L} \right),
    \label{eq:brill_basis0}\\
    B_{l}^1 \left( \theta \right) & = \cos\left( 2l \theta \right).
    \label{eq:brill_basis1}
  \end{eqnarray}
\endnumparts
Here $L$ is a constant that determines the compactification scale and needs to be tuned to the problem being solved. In
our simulations we always use empirically determined value $L=3$ (as usual, in the units of $\sigma$) for the initial data.

Functions $\mathrm{SB}_{k}(x)$ decay as $\frac{1}{x}$ towards infinity, which is the behavior we expect from $\psi$.
Taken together this combination of basis functions automatically satisfies the symmetry and decay properties of the
solution, so we need not impose any explicit boundary conditions. One property this basis set does not guarantee
is regularity of the solution at origin \textemdash{} that would further require that the sub-series corresponding to
each $B_{l}^1$ has a $2l$-th order root at $r=0$ (this is sometimes called the parity theorem \cite{Boyd_2000}). However
it turns out in practice that it is not necessary to impose these conditions explicitly and the solution is regular
anyway.

We implemented the above procedure as a stand-alone library written in C/x86 assembler, which is then called from the
evolution code to construct the initial data. The linear system is solved using LAPACK \cite{lapack_users_guide}.

\paragraph*{Evolution}

Our evolution code is based on the Einstein Toolkit \cite{einsteintoolkit_web,Loffler_2011}, which bundles together the
Cactus framework \cite{cactuscode_web,goodale_2002} and other packages relevant to numerical relativity. The Carpet
\cite{schnetter_2003} code provides Berger-Oliger-style \cite{berger_1984} fixed mesh refinement. The BSSN evolution
equations are implemented by the McLachlan project \cite{mclachlan_web,kranc_web,brown_2008}.

To exploit the axial symmetry of our systems, we use Cartesian coordinates $\left\{ x, y, z \right\}$ with the analytic
Cartoon method, as described in detail in \cite{Hilditch_2016}. Its main idea is replacing the $y$ derivatives of
arbitrary tensors with analytically derived combinations of $x$ and $z$ derivatives. Compared to the \enquote{classical}
Cartoon method \cite{alcubierre_1999} implemented in the Einstein Toolkit, which uses a thin layer of points in the
$y$-direction filled by interpolation, the analytic method has significantly reduced memory requirements and is much
faster. It should also be more accurate, though this was not our main motivation for using it. As our initial data is
also symmetric with respect to reflection through the $z=0$ plane, we evolve just the $z\geq 0$ region.

The simulation domain is thus composed of a set of nested squares. Each one encloses an equidistant grid, with the
step size $\Delta x$ halving per each nesting (refinement) level. We use the method of lines for time evolution, with
4th order Runge-Kutta as the time integrator. Spatial derivatives in the BSSN equations are approximated with 8th order
finite differences \textemdash{} upwind in the advection terms, centered elsewhere. Kreiss-Oliger dissipation of the 9th
order is applied to damp high-frequency noise. When (quasi-)maximal slicing is used, there is also a separate
pseudo-spectral grid present, as described later.

\paragraph*{Maximal slicing}

As the maximal slicing condition \eqref{eq:maximal_slicing} is a linear elliptic PDE, same as the equation
\eqref{eq:brill_ham} we solve for initial data construction, we reuse the basic methods (and some of the code) from
the initial data solver, with a number of changes.

The most obvious of those stem from the fact that the equation is not stand-alone, but a part of a system \textemdash{}
it needs to be solved at each intermediate step of the time integrator and requires the evolved metric variables as
input. Since the evolution happens on the equidistant finite difference grid, we need to interpolate the metric
variables onto the pseudo-spectral grid. Additionally, due to mesh refinement, not all data will always be available at
the required time level, which means we have to interpolate in time as well. We use fourth order Lagrange interpolation
in space and linear in time.

Another change concerns the coordinates. When using polar coordinates we observed that the regularity
conditions at origin stop being satisfied automatically during evolution and the lapse develops a discontinuity at
$r=0$. We believe this happens because the data injected from the finite difference grid is not sufficiently smooth for
the spectral solver. To avoid this issue we use Cartesian coordinates $\left\{ x, z \right\}$ and $B_{k}^0 = B_{k}^1 =
\mathrm{SB}_{2k}$. The scaling constant $L$ is chosen in such a way that the outermost collocation point remains
causally disconnected from the outer boundary during evolution.

Since the equation is now solved many times, performance of the solver becomes important. For this reason we augment
LU decomposition with the BiCGSTAB iterative method \cite{bicgstab}. Since the pseudo-spectral matrix changes slowly
with time, we run the LU decomposition once per $S$ solves and use it to compute the exact inverse. This inverse is then
used as the preconditioner for the next $S$ BiCGSTAB iterative solves. Typically we take $S=1024$, then on average
around 5 iterations are needed for the solution to converge to within $10^{-15}$ absolute error, which takes about
$20\times$ less time to compute than the LU decomposition.

A significant weakness of this approach is that it involves solving a dense linear system of $N^2$ equations, where
$N$ is the number of the basis functions in one dimension. Our memory requirements thus grow as $N^4$, so the maximum
practically achievable number of collocation points is much lower than the number of points on the finite differencing
grid.

With our chosen basis, the collocation points are closely clustered near the origin and become very sparse further out.
So the way lack of resolution manifests in practice is that as the initial wave pulse travels away from the origin, it
enters the area where the distance between collocation points is too large for the waves to be resolved. The higher
frequencies then become invisible to the pseudospectral solver, strongly reducing the accuracy of the solution. Since we
enforce $K=0$ for maximal slicing, the errors manifest as large violations of the ADM constraints. Similar issues have
also been reported in \cite{garfinkle_2001}.

We might be able to mitigate this issue somewhat with a better choice of the basis, or even resolve it fully through
some sort of a multi-domain method. However that would require a significant amount of additional programming and,
since maximal slicing is only an intermediate step of our work, we do not pursue such efforts any further.

\paragraph*{Quasi-maximal slicing}
The obvious approach to computing $W_a$ is reusing the method we used for maximal slicing \textemdash{} solve on each time step,
using time interpolation to fill in missing data points. That, however, consumes a large amount of computational time
and, as previously mentioned, we have considerable freedom to sacrifice accuracy for efficiency in solving
\eqref{eq:qms2}.

\begin{figure}
  \begin{center}
    \includegraphics{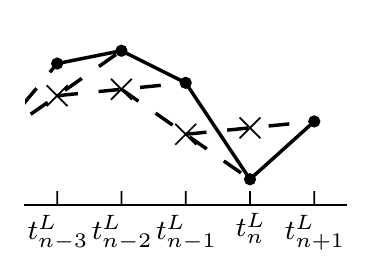}
  \end{center}
  \caption{\label{fig:qms_interp}
    A diagram of the interpolation-extrapolation scheme for evaluating $W_a$. Crosses are the $C_n$s \textemdash{} the
    solutions obtained from the pseudo-spectral solver. Circles are the $C_{n}'$s \textemdash{} the predicted coefficients
    obtained from $C_n$s by extrapolation (dashed lines). Solid lines denote the evaluation of $W_a(t)$ by linear
    interpolation.}
\end{figure}

So we use another method, which requires far less resources and turns out to work well in practice. At its core is the
pseudo-spectral elliptic solver we used for maximal slicing, but it is not run at each time step. It works as follows:
\begin{itemize}
  \item We pick the finest refinement level $L$ that encloses all the collocation points used by the pseudo-spectral solver.
  \item At simulation time $t_{n}^{L}$, before we start the recursive Berger-Oliger evolution to time $t_{n+1}^L$, we
    have the final BSSN quantities for level $L$ and all the finer levels. After interpolating this data onto the
    pseudo-spectral grid, we execute the pseudo-spectral solver to solve \eqref{eq:qms2}, obtaining a set of
    spectral coefficients $C_n$.
  \item During the following recursive time stepping we need the values of $W_a$ at times between $t_{n}^L$ and
    $t_{n+1}^L$, i.e. in the future from the solutions we already have. So we perform {\it linear extrapolation} from
    the last two solutions \textemdash{} $C_{n-1}$ and $C_n$ \textemdash{} to predict the coefficients at time
    $t_{n+1}^L$. We denote this prediction $C_{n+1}'$.
  \item To evaluate $W_a$ at times in the interval $\left(t_{n}^L, t_{n+1}^L\right)$, we {\it linearly interpolate}
    between the two most recent sets of predicted coefficients \textemdash{} $C_{n}'$ and $C_{n+1}'$. This ensures that
    $W_a$ is piecewise linear in time.
\end{itemize}
This process is sketched in Figure \ref{fig:qms_interp}.

Finally, we discovered empirically that simply evaluating $W_a$ as a sum of the spectral series tends to introduce
strong high-frequency components to the finite difference grid. It seems that those consequently propagate to $\alpha$
and from there to the other evolved functions and due to non-linearities get aliased as spurious low-frequency noise,
manifesting as ADM constraint violations.

We deal with this issue by applying a low-pass filter to the basis functions. In our current multi-grid setup, the
Nyquist frequency at a given grid point depends on the distance from origin $r$. For each basis function $B_{k}(x)$ we
estimate its \enquote{base} frequency from the location of its first zero $x_{k}^0$, which implies it is no longer
accurately resolved at the radius $r_{k}^0$ where $\Delta x \approx x_{k}^0$. Therefore when evaluating $W_a$, we damp each
basis function such that \cite{szilagyi_2009}
\begin{equation}
  B_{k}' = B_{k} \mathrm{exp} \left( - J \left(\frac{r}{r_{k}^0}\right)^d \right).
  \label{eq:qms_damping}
\end{equation}
$J=36$ ensures that for $r=r_0$ the exponential evaluates to the double-precision $\varepsilon$, while
the empirically determined choice of $d=6$ makes sure the transition area is never smaller than the roughest step size
(the results do not seem to depend on the precise value of $d$ very strongly). Altogether the filter ensures that $W_a$
remains non-zero only close to the origin and does not affect the grid further out, where it is not needed.

At late simulation times we usually switch from quasi-maximal back to 1+log slicing, since the added term no longer
seems to be necessary for stability and for super-critical spacetimes the spectral solver can be ill-behaved in the
presence of punctures. The switch is implemented by using a damping function similar to \eqref{eq:qms_damping} for
$\kappa$ in \eqref{eq:qms}, specifically
\begin{equation}
  \kappa = \mathrm{min}\left[\mathrm{exp}\left( -J \left( t - t_0 \right)^d \right), 1\right],
  \label{eq:qms_switchoff}
\end{equation}
where $t_0$ is the time at which the switch starts, determined empirically for a given simulation. The exact choice of
parameters did not make a qualitative difference in our runs, $t_0$ is rougly the time when the apparent horizon forms
for super-critical runs or the time when the largest peak in $\mathcal{K}$ forms and dissolves.
\subsection{Code validation}
\label{sec:validation}

\paragraph*{Super-critical Brill waves in moving puncture gauge}

\begin{figure}
  \begin{minipage}[b]{0.45\textwidth}
    \includegraphics{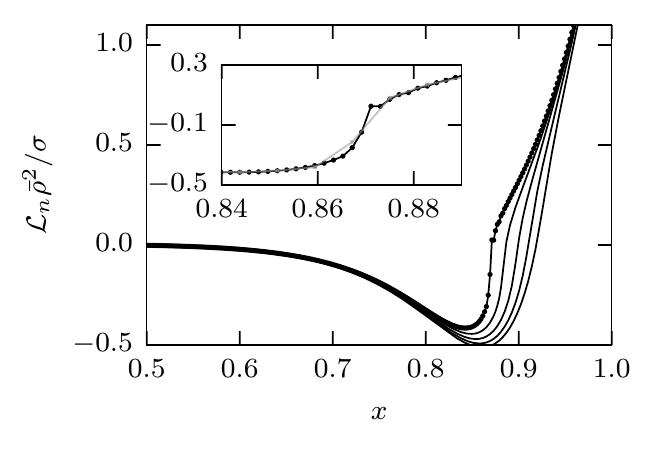}
  \end{minipage}
  \begin{minipage}[b]{0.45\textwidth}
      \includegraphics{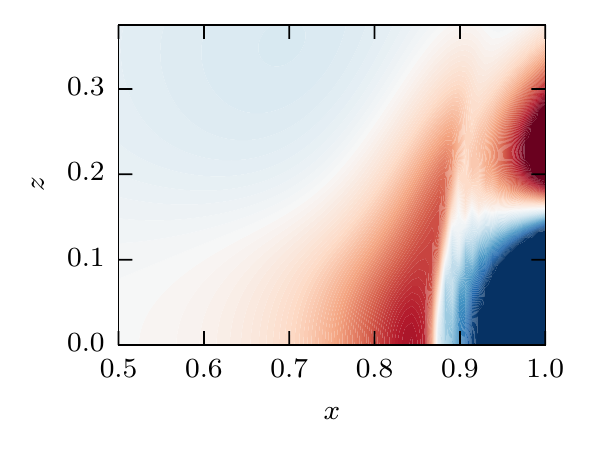}
  \end{minipage}
  \begin{minipage}[b]{0.05\textwidth}
      \includegraphics{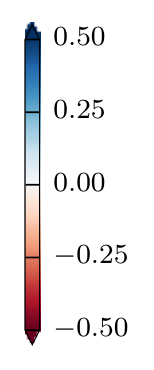}
      \vspace{1.1em}
  \end{minipage}
  \caption{\label{fig:mp_a5_rho}
    Failure of the 1+log slicing for $A=5$ Brill waves. The shift is set to zero. Left: plotted is the derivative
    of $\bar{\rho}^2$ along the normal to the spatial surfaces on the $x$-axis at times $\left\{ 4.625, 4.6875, 4.75,
    4.8125, 4.875 \right\}$ (bottom to top). The inset compares the data on the $t=4.875$ slice for runs with 2048
    (black) and 512 (gray) grid points. Circles/crosses indicate the positions of the grid points. Right: the same
    quantity in the $x-z$ plane at time $t=4.875$.}
\end{figure}

We validate our moving puncture gauge setup by reproducing the slicing failure for $A=5$ Brill waves from
\cite{Hilditch_2013}. We use 6 levels of mesh refinement, with $2048\times 2048$ grid points on each, the coarsest level
using a spatial step size of $\Delta x = \frac{1}{4}$. The $\Gamma$-driver parameters are set to $\xi=1; \eta = 11.4 / \sigma$.
Our results are fully consistent with the cited paper.

The slicing issues are illustrated in Fig. \ref{fig:mp_a5_rho}.  The circumferential radius $\bar{\rho}$ should be a
smooth spacetime invariant, so for a regular hypersurface $t=\mathrm{ const.}$ the derivative of $\bar{\rho}$ along the
unit normal $n^\mu$ should also be smooth. In the left panel we show that at times $t \lesssim 4.9$ \textemdash{} just
before the simulations fail (independently of the spatial resolution) \textemdash{} this function develops a feature
resembling a discontinuity. As can be seen in the inset, the gradient becomes higher with increasing number of grid
points. When testing with a number of different resolutions from 384 to 2048 grid points, we found that the maximum of
the gradient of the plotted function exhibits non-convergent behavior consistent with computing finite differences
across a step function. We thus conjecture that the spatial slices themselves become non-smooth, which leads to
simulation failure. The right panel then shows the \enquote{break} in the same quantity in the $x-z$ plane.

\paragraph*{Quasi-maximal slicing}

\begin{figure}
  \begin{minipage}{0.5\textwidth}
    \includegraphics{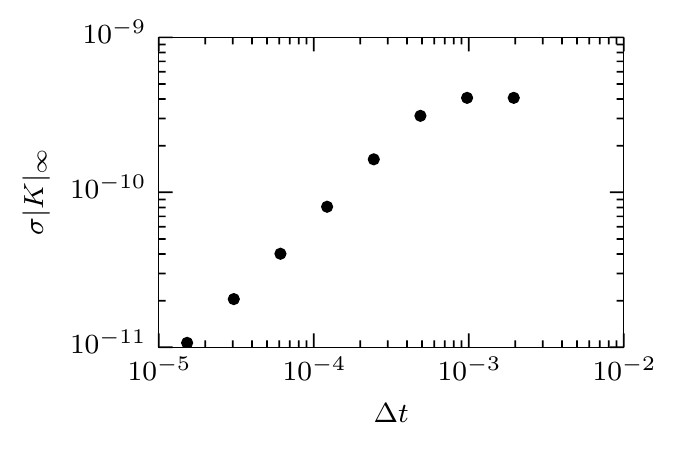}
  \end{minipage}
  \begin{minipage}{0.5\textwidth}
    \includegraphics{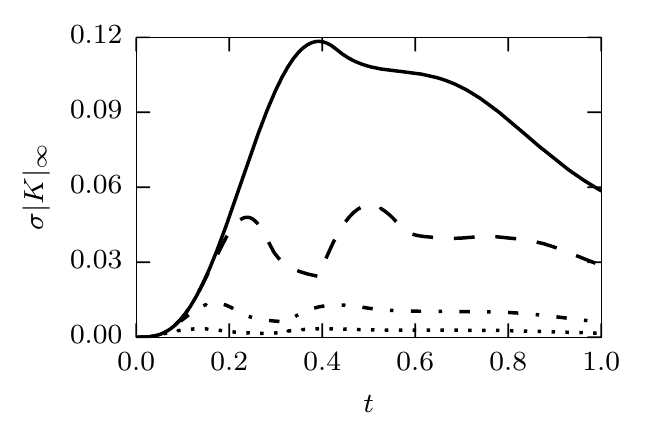}
  \end{minipage}
  \caption{\label{fig:qms_convergence}
    Convergence of the quasi-maximal slicing with the time step $\Delta t$. Left: $A=0.1$ Brill waves, plotted
    is the maximum norm of $K$ at simulation time $t=\frac{1}{512}$ (one full time step for the coarsest run) as a function
    of the time step $\Delta t$. Right: $A=1$ Brill waves, time evolution of the maximum norm of $K$ over the finest
    refinement level for (top to bottom): 1+log slicing and quasi-maximal slicing with $\Delta x / \Delta t =
    \left\{ \frac{1}{4}, \frac{1}{8}, \frac{1}{16} \right\}$ (changing the CFL factor for the 1+log run has no visible
    effect on the plot within the bounds of stability). See main text for more details.}
\end{figure}

\begin{figure}
  \begin{minipage}[t]{0.45\linewidth}
    \includegraphics{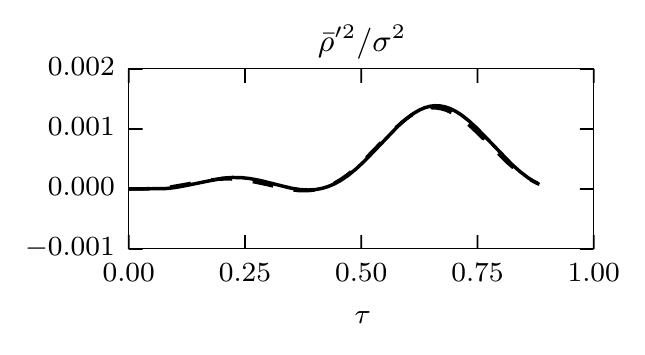}\\
  \end{minipage}
  \begin{minipage}[t]{0.45\linewidth}
    \includegraphics{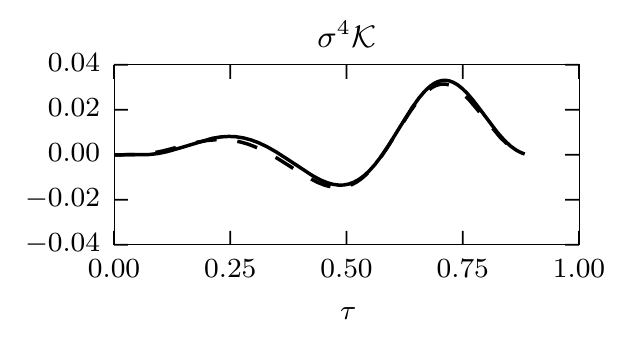}
  \end{minipage}
  \caption{\label{fig:qms_convergence_evol}
    Convergence of spacetime invariants with quasi-maximal slicing for $A=1$ Brill waves. Invariant quantities
    at $\bar{\rho} = 1$ \textemdash{} left: $\bar{\rho}'^2$, right: $\mathcal{K}$ \textemdash{} are computed as functions of
    the proper time $\tau$ at a given location for simulations with $\left\{ 256, 384, 512 \right\}$ grid points per refinement
    level. Plotted is the difference between the $256-384$-point run (solid line) and the $384-512$-point run (dashed
    line), scaled for 4th order convergence.}
\end{figure}

First, we verify our solver for the equation \eqref{eq:qms2} does indeed converge to $\partial_{tt}K = 0$. To that end,
we run a series of simulations with weak waves ($A = 0.1$) and no mesh refinement, where $W$ is evaluated at each
full step of the time integrator (i.e. not using the extrapolation/interpolation procedure from Section
\ref{ssec:implementation}). The simulations differ only in the time step $\Delta t$, which halves for each finer run
\textemdash{} the coarsest run makes just one full time step, while the finest one makes 128. As the left panel of
Figure \ref{fig:qms_convergence} shows, the maximum norm of $K$ goes linearly to zero as the time step decreases.

Next, we run a series of simulations for stronger waves ($A=1$), with the grid setup as for the previously described
1+log runs.  The reference run uses the 1+log slicing, and we compare it against several quasi-maximally sliced runs
(now as described in Section \ref{ssec:implementation}) with varying CFL factor. As can be seen from the top panel of
Figure \ref{fig:qms_convergence}, even when $W_a$ is a very crude approximation to the solution of $\eqref{eq:qms2}$,
$K$ is significantly lower for quasi-maximal slicing than for 1+log, and further decreases with smaller time step.

The vanishing of $K$ as the time step goes down is slow and hard to quantify, so we find it more useful to regard
quasi-maximal slicing merely as a procedure that gives us some slicing which is \enquote{close} to the maximal one, but
distinct from it. What we want to verify then is that when we hold the parameters of the quasi-maximal solver fixed and
increase the resolution of the finite difference grid, the simulations converge to the same
underlying spacetime, though possibly in different slicing. For that purpose, we look at spacetime invariants.

We run three simulations, again with $A=1$, and $\left\{ 256, 385, 512 \right\}$ grid points per refinement level. For
this data, the circumferential radius $\bar{\rho}$ in the equatorial plane $z=0$ remains a monotonous function of $x$
throughout the simulation, so it can be used as a radial coordinate (this is not the case for stronger Brill waves). For
each simulation we calculate the value of $\bar{\rho}'^2$ and $\mathcal{K}$ seen by the observer at $\bar{\rho}=1$. To
work with purely invariant quantities, we also compute that observer's proper time $\tau$ by integrating
$\mathrm{d}\tau = -(g_{\mu\nu} \mathrm{d}x^\mu \mathrm{d}x^\nu)^\frac{1}{2}$ and plot the aforementioned invariants as
functions of $\tau$. As we can see from Figure \ref{fig:qms_convergence_evol}, those values converge to
each other with the fourth order, which is the order of the time integrator. We can thus conclude that quasi-maximally
sliced simulations (with fixed pseudo-spectral solver order) do indeed converge to the same physical spacetime. As
previously mentioned, it would be hard to state precisely the measure of convergence towards the maximal slicing.
\section{Evolution of Brill waves with quasi-maximal slicing}
\label{sec:results}

\paragraph{Near-critical case: $A=5$}

\begin{figure}[hbt]
  \begin{minipage}{0.45\linewidth}
    \includegraphics{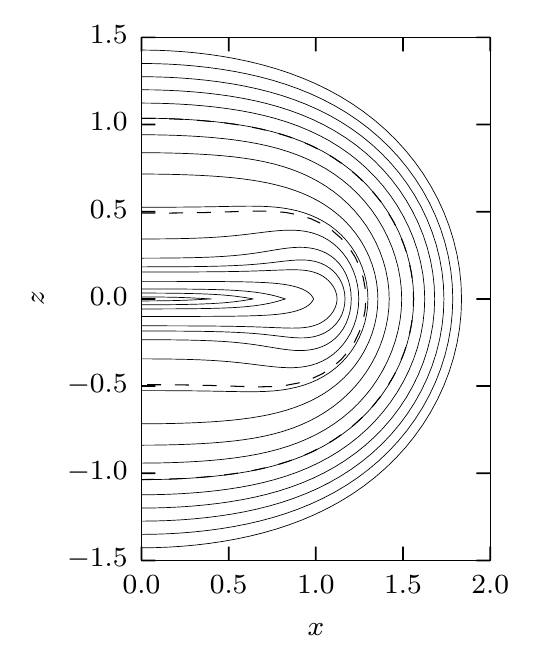}
  \end{minipage}
  \begin{minipage}{0.45\linewidth}
    \includegraphics[scale=0.3]{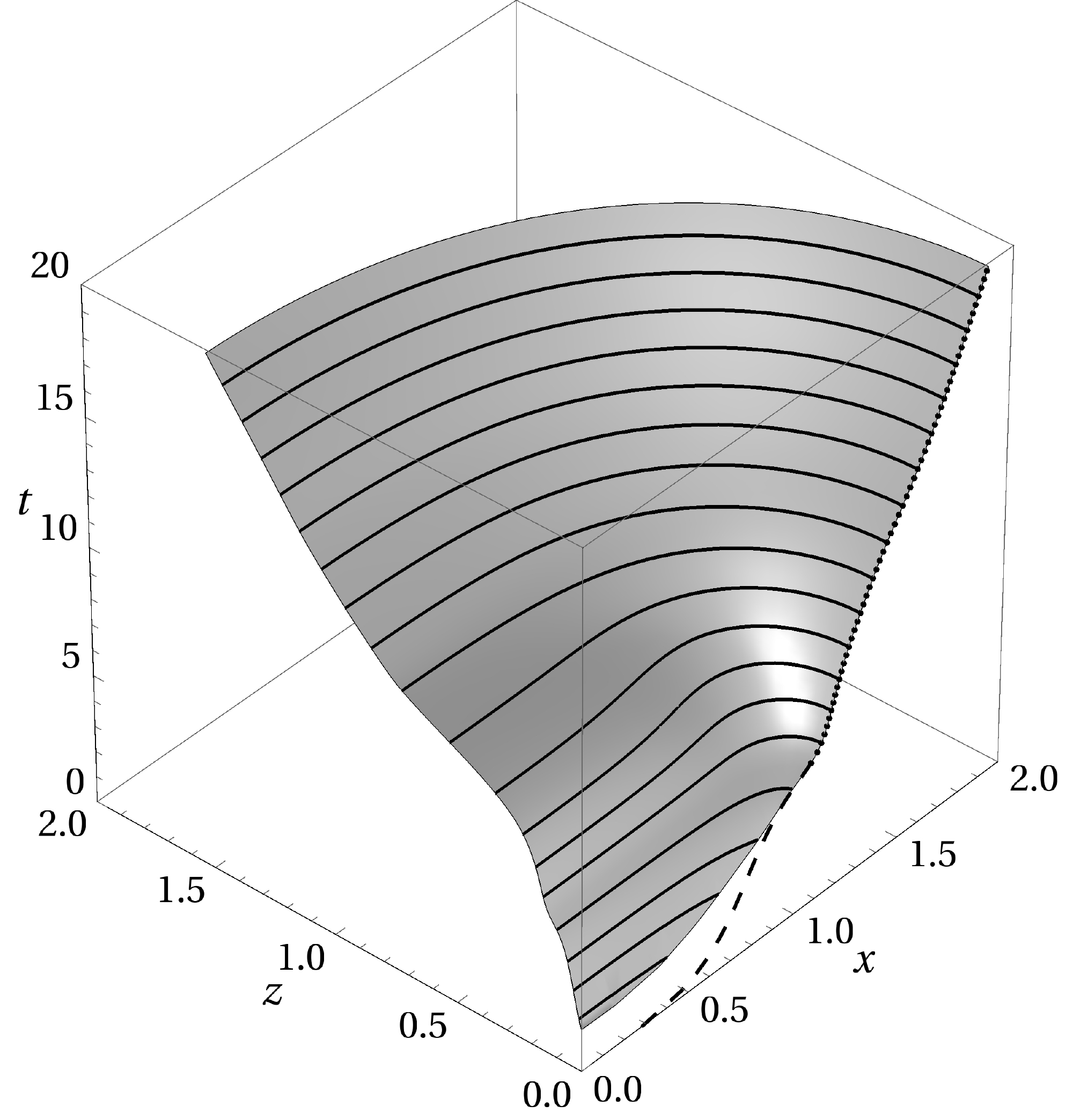}
  \end{minipage}
  \caption{\label{fig:hor_evol}
    Formation of the event horizon for $A=5$ Brill waves, zero shift. It is constructed by evolving the apparent horizon
    section from the $t=25$ slice as a null surface back in time.
    Left: solid lines show cross-sections of the event horizon in
    the $x-z$ plane at simulation times $t=\left\{ 2, 3, \cdots, 14 \right\}$ (inner to outer). Dashed lines show the
    apparent horizon at time $t=10$ (when it first appears) and $t=14$. At later times both horizons cannot be
    distinguished in this plot. Right: the event horizon as a surface in $x-z-t$ coordinates.  When it first appears it
    is not smooth at the equator, which is represented by the $x$ coordinate in this plot. It remains non-smooth until
    the radial null geodesic (dashed line in the $z=0$ plane) enters the event horizon as its generator (dotted line).}
\end{figure}

\begin{figure}[hbt]
  \begin{minipage}{0.47\linewidth}
    \includegraphics{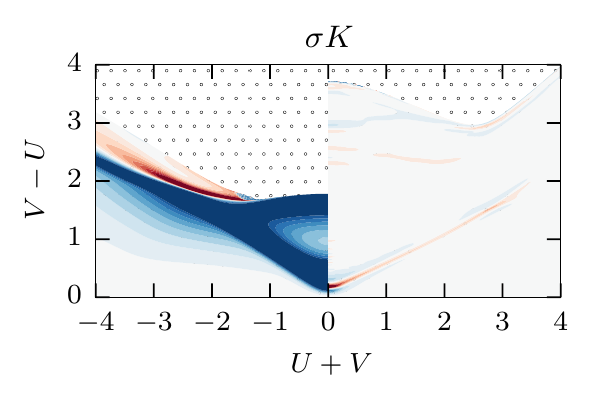}\\
    \includegraphics{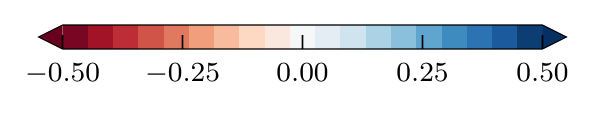}
  \end{minipage}
  \begin{minipage}{0.47\linewidth}
    \includegraphics{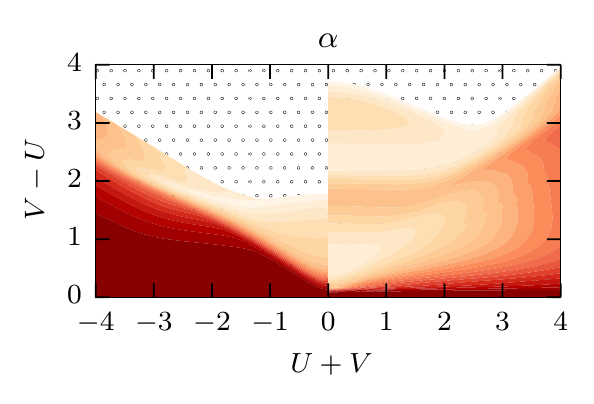}\\
    \includegraphics{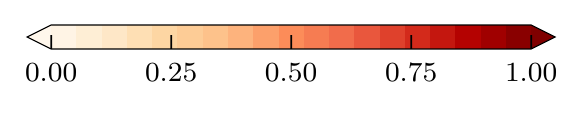}
  \end{minipage}
  \caption{\label{fig:a5_doublenull}
    Comparison of the $A=5$ runs with 1+log / quasi-maximal slicings \textemdash{} respectively the left and
    right half of each panel \textemdash{} in the double-null coordinates $\left\{ U, V \right\}$ in the equatorial
    plane. Color denotes the value of $K$ (left panel) and $\alpha$ (right panel), dotted regions are not covered by
    respective simulations.}
\end{figure}

Our main result is demonstrating that quasi-maximal slicing cures the singularities arising with 1+log slicing for $A=5$
Brill waves. This near-critical initial data, with ADM mass $0.70$, was used as the failing case in
\cite{Hilditch_2013}. Though quasi-maximal slicing resolves the singularities from the $\alpha$/$K$ sector, we have
encountered additional issues with the shift.  When using the $\Gamma$-driver, the shift seems to develop a shock
profile around $t\approx 10$. For that reason, we use zero shift in the simulations described here. While that allows us
to run the simulation further, it is well known that a non-zero shift vector is required in black-hole spacetimes to
avoid \enquote{slice stretching} \textemdash{} we are thus still unable to evolve this data indefinitely.

Around simulation time $t\approx10.25$ we first find an apparent horizon with mass $M_{AH} = 0.56$, which grows to
$M_{AH}=0.58$ by $t=20$. The masses agree with values reported in \cite{hilditch_2017}. By integrating the apparent
horizon to the past as a null surface we construct the event horizon, seen in Figure \ref{fig:hor_evol}. Its shape at
early times turns out to be that of a disk with a non-smooth rim.  This can be understood from the curious fact that if
we shoot photons from the origin at $t=0$, those along the $z$-axis will escape to infinity, while those along the
$x$-axis will end up below the horizon. The event horizon thus does not intersect the initial slice and appears around
$t=1.7$, when the first photon in the $z$ direction fails to escape. Its worldline along the $x$-axis then has a spatial
character until it connects with its generating null geodesic, as is sketched in the right panel of Figure
\ref{fig:hor_evol}. This means the event horizon can briefly exist as a torus, which we observed for a run with
quasi-maximal slicing and $A=4.8$.

There are various sources of uncertainty in determination of the event horizon. The most important one is the error of
the simulation, which can be judged by relative variation of late-time MOTS areas between various simulations and
appears to be of order $10^{-3}$. Of similar order is the error with which we can trace null geodesics forming the event
horizon (this is more accurate along the $x$ and $z$ axes, where the data available to the tracing code is much better
resolved in time). Then there is the difference between the event horizon and MOTS and the uncertainty of future
evolution of the spacetime, but due to unstable nature of future-oriented null geodesics near the event horizon, we find
that at $t=10$ a spatial variation of $\Delta x \approx 10^{-5}$ will \textemdash{} during the runtime of the simulation
\textemdash{} either direct the null ray inside the black hole or put it on an escape trajectory.

To demonstrate that the quasi-maximal run actually does cover a larger chunk of spacetime than the crashing 1+log run
(rather than merely collapsing the lapse in the critical area), we compare the two slicings in double-null coordinates
$\left\{ U, V \right\}$.  Those are constructed by integrating null geodesics along the $x$-axis in, respectively, the
positive and negative $x$ directions. The values of $U$ and $V$ in the initial slice $t=0$ are both set equal to half
the proper distance from the origin. Figure \ref{fig:a5_doublenull} clearly shows that quasi-maximal slicing advances
farther towards the physical singularity in the central area.

\paragraph{Super-critical case: $A=5.5$}

\begin{figure}
  \begin{minipage}{0.5\textwidth}
    \includegraphics{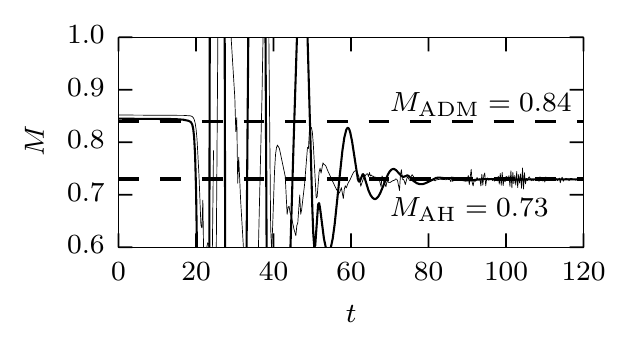}
  \end{minipage}
  \begin{minipage}{0.5\textwidth}
    \includegraphics{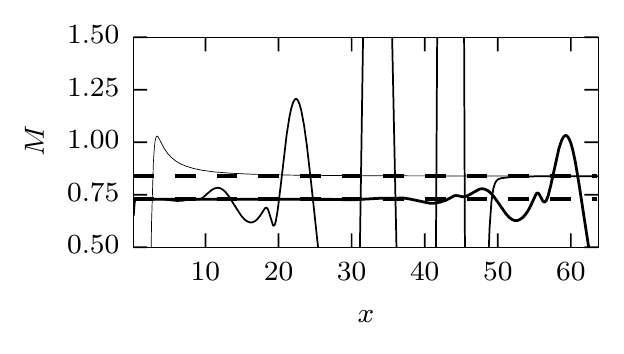}
  \end{minipage}
  \caption{\label{fig:mass_evol}
    Evolution of mass estimates for $A=5.5$ Brill waves. Left: evolution of $M_\mathcal{K}$ (thin line) and
    $M_{\bar{\rho}}$ (thick line) at constant spatial coordinate $\left\{ x = 20, z = 0 \right\}$ with the simulation
    time $t$. Right: $M_{\bar{\rho}}$ along the $x$-axis at times $t = \left\{ 0, 50, 100 \right\}$ (thin to thick). In
    both pictures the dashed horizontal lines indicate the ADM mass of the initial slice (upper) and the final apparent horizon
    mass $M_{AH}$ (lower). See the main text for details.}
\end{figure}

The ADM mass of this data is $0.84$. Unlike the previous case, we are able to use the $\Gamma$-driver without any
pathologies, which allows the simulations to run much longer. At $t\approx 7.5$ we find an apparent horizon with mass
$M_{AH} = 0.63$, which quickly increases to its final value of $0.73$.

At later times, the spacetime should evolve into a central region that rapidly relaxes into a Schwarzschild black hole,
plus an outward-travelling wave packet. We illustrate this fact in Figure \ref{fig:mass_evol} by plotting mass estimates
\eqref{eq:mass_kretschmann} and \eqref{eq:mrho}. These, as mentioned, have the meaning of mass only in the static
end-state, so we can see them oscillate wildly as the waves radiate away. At late simulation times they closely approach
the value of $M_{AH}$, which shows that we really obtain a Schwarzschild black hole with expected properties.

The $\Gamma$-driver behaves in a manner similar to that known for the evolutions of Schwarzschild starting with
\enquote{wormhole} slices, effectively excising the central region from the numerical grid.  At late times the numerical
slices terminate at $\bar{\rho} \approx 1.31$.

\paragraph{Other near-critical amplitudes}

\begin{figure}
  \begin{minipage}{0.45\linewidth}
    \includegraphics{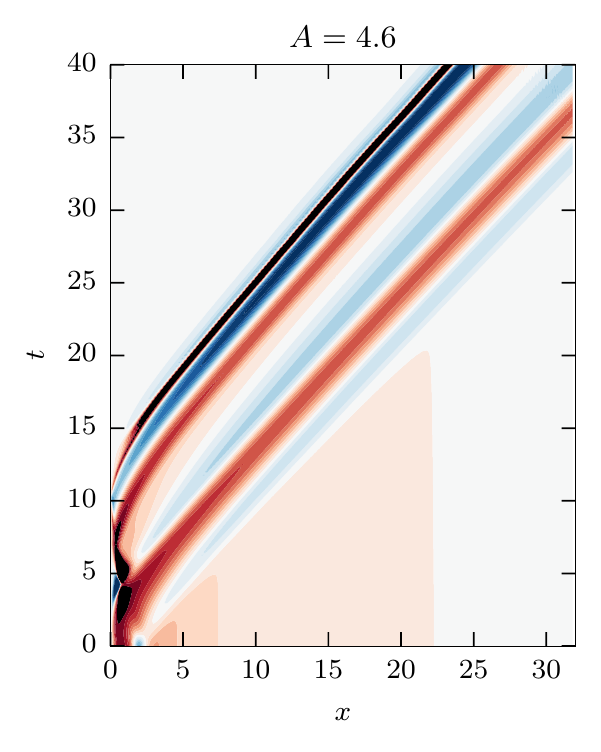}
  \end{minipage}
  \begin{minipage}{0.45\linewidth}
    \includegraphics{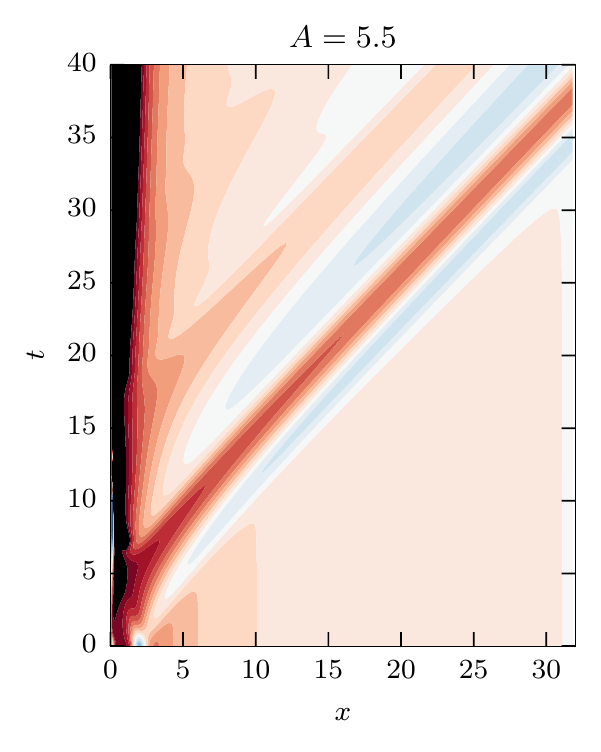}
  \end{minipage}\\
  \begin{center}
    \includegraphics{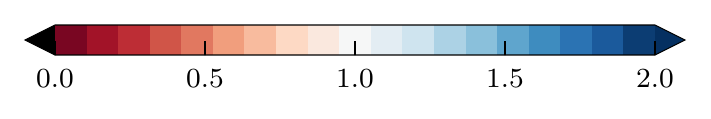}
  \end{center}
  \caption{\label{fig:rho_evol}
    Evolution of $\chi$ in the equatorial plane for Brill waves with $A=4.6$ (left) and $A=5.5$ (right).}
\end{figure}

\begin{figure}
  \begin{minipage}[b]{0.45\linewidth}
    \includegraphics{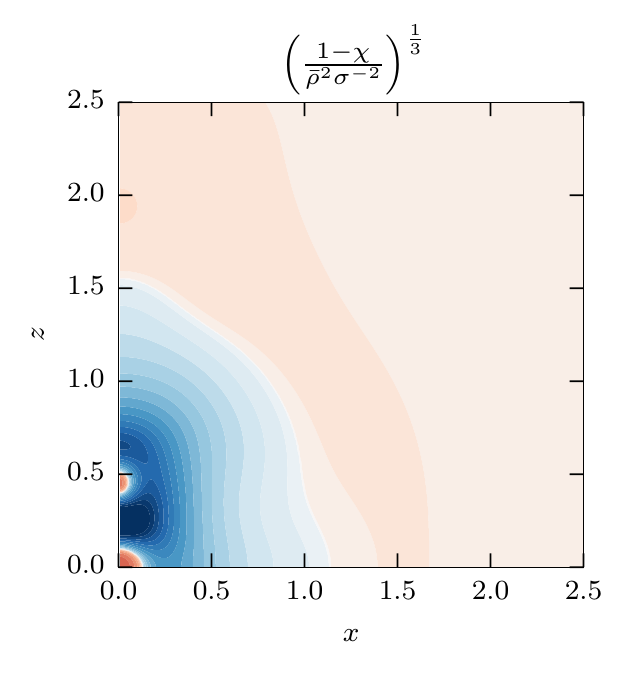}\\
    \includegraphics{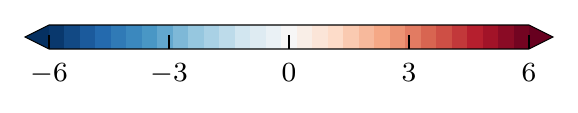}
  \end{minipage}
  \begin{minipage}[b]{0.45\linewidth}
    \includegraphics{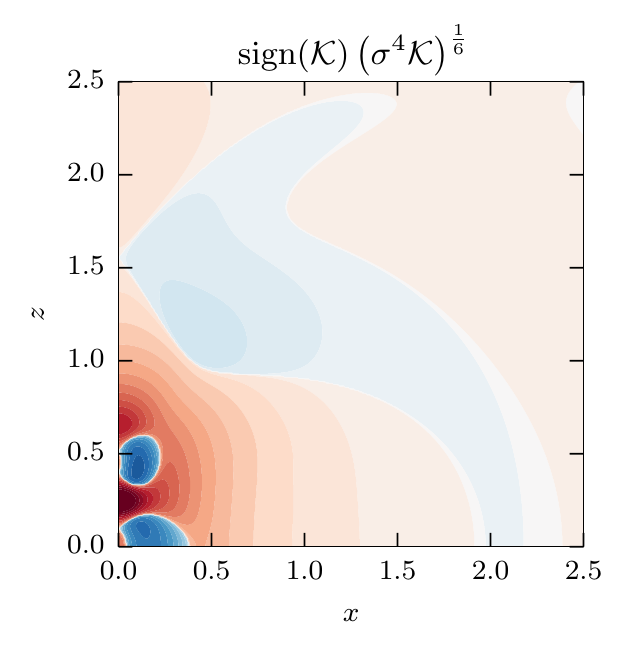}\\
    \includegraphics{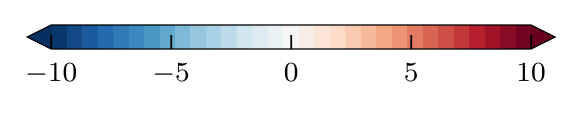}
  \end{minipage}\\
  \caption{\label{fig:invar_consttime}
Invariants in the $x-z$ plane at coordinate time when they reach their highest values ($t=11$); $A=4.65$ data, zero shift.}
\end{figure}

\begin{figure}
  \centering
  \begin{minipage}[t]{0.45\textwidth}
    \includegraphics{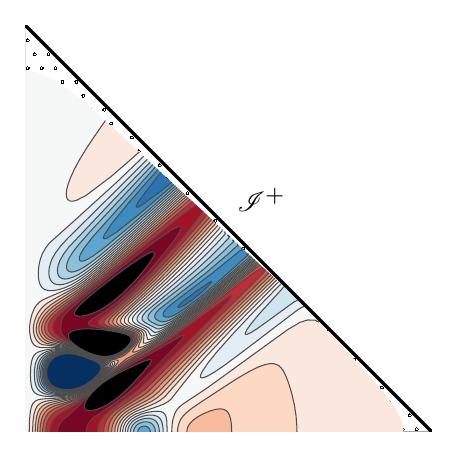}
  \end{minipage}
  \begin{minipage}[t]{0.45\textwidth}
    \includegraphics{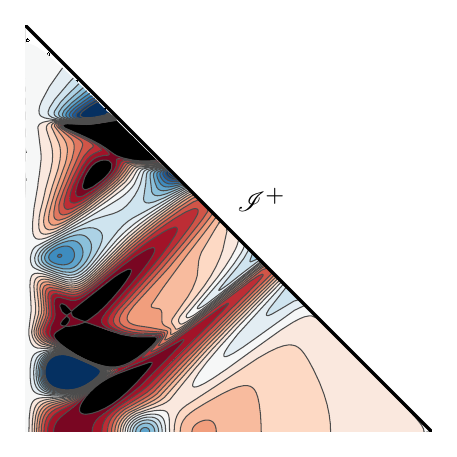}
  \end{minipage}\\

  \begin{minipage}[t]{0.45\linewidth}
    \includegraphics{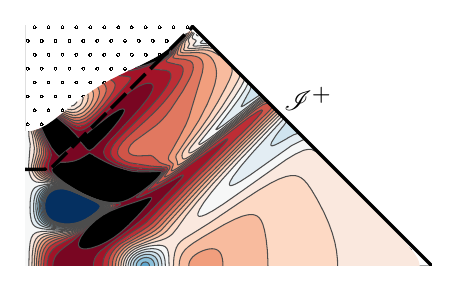}
  \end{minipage}
  \begin{minipage}[t]{0.45\linewidth}
    \includegraphics{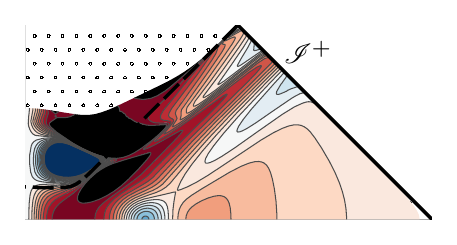}
  \end{minipage}

  \vspace{1em}
  \includegraphics{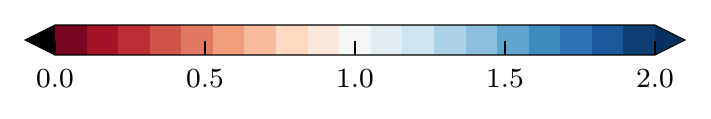}

  \caption{\label{fig:brill_conformal}
    Conformal diagrams for the equatorial plane of the spacetimes produced by evolving the Brill waves. The
    values of $A$ are $\left\{ 4.0, 4.65, 4.8, 5.0 \right\}$ for, respectively, top left, top right, bottom left, and
    bottom right. The top two spacetimes are sub-critical, while the bottom two are super-critical, as seen from the
    presence of an event horizon (dashed lines). The spacelike portion of the event horizon is non-smooth, as
    described in more detail in the main text. The color denotes the value of $\chi$, defined by \eqref{eq:chi}. Note
    the range the quantity $\chi$ spans \textemdash{} for the spherically symmetric sub-critical spacetimes of
    \cite{Choptuik_1993} it stays between $0.4$ and $1$.}
\end{figure}

We run a number of additional simulations with both sub- and super-critical initial data until, respectively, the waves
disperse in the central region or an apparent horizon forms. We observe no slicing singularities in any of them.

For analyzing the spacetime dynamics we rely on the invariants introduced in Section \ref{sec:invariants}. Figure
\ref{fig:rho_evol} shows values of the scalar $\chi$, defined by \eqref{eq:chi}, for long-term evolutions of sub-
and super-critical waves. As previously mentioned, $\chi$ is negative when the gradient of the circumferential radius is
timelike. In Schwarzschild this is true inside the event horizon, with $\nabla^\mu R$ pointing to the future in the
\enquote{white hole} region and to the past in the \enquote{black hole} region. It is interesting that for Brill waves,
the orientation is to the future in the first negative-$\chi$ region (i.e. \enquote{the wrong way}) and then switches
direction, possibly multiple times, which shows how wildly dynamic those spacetimes are.

A drawback of $\chi$ is that it only has a clear mapping to Schwarzschild in the equatorial plane, since physically
meaningful values of $\theta$ are not available. One way around that is the scalar $\frac{1 - \chi}{\bar{\rho}^2}$,
which is equal to $\frac{2M}{R^3}$ in Schwarzschild. It is, however, no longer dimensionless. This quantity is shown
alongside $\mathcal{K}$ in Figure \ref{fig:invar_consttime} for a sub-critical run at the time when those scalars attain
its highest values. Interestingly this happens on the $z$ axis some distance away from the origin (cf. the results of
\cite{hilditch_2017}).

To compare the spacetime geometries with differing amplitude parameter $A$ side-by-side, we mimic the usual
compactification of the Schwarzschild manifold by a transformation that maps the positive part of the $x$-axis through
$X = \mathrm{arctan}\left[ \frac{3}{2}\mathrm{sinh}\left(\frac{x}{6}\right) \right]$ into the interval $(0,
\frac{\pi}{2})$ and then maps the null geodesics in the $x-t$ plane into straight $45$-degree lines. This produces
the conformal diagrams in Figure \ref{fig:brill_conformal}, which illustrate propagation of waves towards future null
infinity.
\section{Summary}
\label{sec:summary}

We run numerical simulations of sub- and super-critical Brill wave initial data. We manage to verify previously
described slicing problems for strong data with a different code and illustrate them with figures showing that the
$t=\mathrm{const.}$ slices develop a cusp-like shape. These problems turn out to be removable by amending the 1+log
slicing with a source term derived from the maximal slicing condition.

This new slicing condition, which we call the \enquote{quasi-maximal} slicing, improves the regularity of evolved
spatial slices by pushing them closer towards the maximal ones. We describe long-term evolution of the spacetimes
arising from near-critical waves and construct their conformal diagrams. Of special note is the non-smooth shape of
the event horizon for weaker data. This data is not obtainable with the 1+log slicing because of the aforementioned
pathologies.

As mentioned in the introduction, Hilditch et al.\ are currently working on the same problem with a new pseudospectral
code using a generalized harmonic formulation \cite{Hilditch_2016,hilditch_2017}, so a brief comparison is in order.
Since our code is based on finite differences and the moving puncture approach, it inherits their advantages, such as
simplicity, robustness and the ability to run long-term simulations containing black holes without the need for
excision. The price to pay is mainly lower efficiency, i.e. higher resolution required for the same accuracy.  Of
course, the fact that both these approaches produce the same results (as confirmed in private communication), is greatly
encouraging.

The code used for running the simulations described in this paper is available for download through the \textit{git}
protocol at \url{git://git.khirnov.net/qms_source_2017}.

\ack
We would like to thank David Hilditch for very productive discussions.  This work is supported by the Charles
University in Prague, project GA UK No 2000314, GA UK No 1176217 and SVV-260211. T.L. acknowledges the support from the
Czech Science Foundation grant No 14-37086G (A. Einstein Center). Computational resources were provided by the CESNET
LM2015042 and the CERIT Scientific Cloud LM2015085, provided under the programme \enquote{Projects of Large Research,
Development, and Innovations Infrastructures}. We would also like to thank the developers of the Einstein Toolkit for
making open-source numerical relativity possible.

\section*{References}

\begin{thebibliography}{10}
\providecommand{\url}[1]{\texttt{#1}}
\providecommand{\urlprefix}{URL }
\providecommand{\eprint}[2][]{\url{#2}}

\bibitem{eppley_1977}
Eppley K 1977 Evolution of time-symmetric gravitational waves: Initial data and
  apparent horizons \emph{Phys. Rev. D} \textbf{16} \href{https://link.aps.org/doi/10.1103/PhysRevD.16.1609}{1609--1614}
  

\bibitem{Brill_1959}
Brill D~R 1959 On the positive definite mass of the bondi-weber-wheeler
  time-symmetric gravitational waves \emph{Annals of Physics} \textbf{7}(4) 
  \href{http://www.sciencedirect.com/science/article/pii/0003491659900557}{466-- 483}

\bibitem{Abrahams_1993}
Abrahams A~M and Evans C~R 1993 Critical behavior and scaling in vacuum
  axisymmetric gravitational collapse \emph{Phys. Rev. Lett.} \textbf{70}
  \href{http://link.aps.org/doi/10.1103/PhysRevLett.70.2980}{2980--2983}

\bibitem{Teukolsky_1982}
Teukolsky S~A 1982 Linearized quadrupole waves in general relativity and the
  motion of test particles \emph{Phys. Rev. D} \textbf{26} 
  \href{http://link.aps.org/doi/10.1103/PhysRevD.26.745}{745--750}

\bibitem{Choptuik_1993}
Choptuik M~W 1993 Universality and scaling in gravitational collapse of a
  massless scalar field \emph{Phys. Rev. Lett.} \textbf{70} 
  \href{http://link.aps.org/doi/10.1103/PhysRevLett.70.9}{9--12}

\bibitem{gundlach_2007}
Gundlach C and Mart{\'i}n-Garc{\'i}a J~M 2007 Critical phenomena in
  gravitational collapse \emph{Living Reviews in Relativity} 
  \href{http://dx.doi.org/10.12942/lrr-2007-5}{\textbf{10}(1) 5}

\bibitem{alcubierre_2000}
Alcubierre M, Allen G, Br\"ugmann B, Lanfermann G, Seidel E, Suen W~M and
  Tobias M 2000 Gravitational collapse of gravitational waves in 3d numerical
  relativity \emph{Phys. Rev. D} \textbf{61} 
  \href{http://link.aps.org/doi/10.1103/PhysRevD.61.041501}{041501}

\bibitem{garfinkle_2001}
Garfinkle D and Duncan G~C 2001 Numerical evolution of brill waves \emph{Phys.
  Rev. D} \textbf{63} 
  \href{https://link.aps.org/doi/10.1103/PhysRevD.63.044011}{044011}

\bibitem{bona_1995}
Bona C, Mass\'o J, Seidel E and Stela J 1995 New formalism for numerical
  relativity \emph{Phys. Rev. Lett.} \textbf{75}
  \href{https://link.aps.org/doi/10.1103/PhysRevLett.75.600}{600--603}

\bibitem{alcubierre_2003}
Alcubierre M, Br\"ugmann B, Diener P, Koppitz M, Pollney D, Seidel E and
  Takahashi R 2003 Gauge conditions for long-term numerical black hole
  evolutions without excision \emph{Phys. Rev. D} \textbf{67} 
  \href{https://link.aps.org/doi/10.1103/PhysRevD.67.084023}{084023}

\bibitem{Hilditch_2013}
Hilditch D, Baumgarte T~W, Weyhausen A, Dietrich T, Br\"ugmann B, Montero P~J
  and M\"uller E 2013 Collapse of nonlinear gravitational waves in
  moving-puncture coordinates \emph{Phys. Rev. D} \textbf{88} 
  \href{http://link.aps.org/doi/10.1103/PhysRevD.88.103009}{103009}

\bibitem{Hilditch_2016}
Hilditch D, Weyhausen A and Br\"ugmann B 2016 Pseudospectral method for
  gravitational wave collapse \emph{Phys. Rev. D} \textbf{93} 
  \href{http://link.aps.org/doi/10.1103/PhysRevD.93.063006}{063006}

\bibitem{hilditch_2017}
Hilditch D, Weyhausen A and Br\"ugmann B 2017 Evolutions of centered brill
  waves with a pseudospectral method \emph{Phys. Rev. D} \textbf{96} 
  \href{https://link.aps.org/doi/10.1103/PhysRevD.96.104051}{104051}

\bibitem{rinne_2008}
Rinne O 2008 Constrained evolution in axisymmetry and the gravitational
  collapse of prolate brill waves \emph{Classical and Quantum Gravity}
  \textbf{25}(13) 
  \href{https://iopscience.iop.org/article/10.1088/0264-9381/25/13/135009}{135009}

\bibitem{Shibata_1995}
Shibata M and Nakamura T 1995 Evolution of three-dimensional gravitational
  waves: Harmonic slicing case \emph{Phys. Rev. D} \textbf{52} 
  \href{http://link.aps.org/doi/10.1103/PhysRevD.52.5428}{5428--5444}

\bibitem{Baumgarte_1998}
Baumgarte T~W and Shapiro S~L 1999 On the numerical integration of einstein's
  field equations \emph{Phys. Rev. D} \textbf{59} 
  \href{http://link.aps.org/doi/10.1103/PhysRevD.59.024007}{024007}

\bibitem{alcubierre_2008}
{Alcubierre} M 2008 \emph{{Introduction to 3+1 Numerical Relativity}} (Oxford
  University Press, UK) ISBN 978-0-19-920567-7

\bibitem{Gundlach_2006}
Gundlach C and Mart\'{\i}n-Garc\'{\i}a J~M 2006 Well-posedness of formulations
  of the einstein equations with dynamical lapse and shift conditions
  \emph{Phys. Rev. D} \textbf{74} 
  \href{http://link.aps.org/doi/10.1103/PhysRevD.74.024016}{024016}

\bibitem{van_meter_2006}
van Meter J~R, Baker J~G, Koppitz M and Choi D~I 2006 How to move a black hole
  without excision: Gauge conditions for the numerical evolution of a moving
  puncture \emph{Phys. Rev. D} \textbf{73} 
  \href{https://link.aps.org/doi/10.1103/PhysRevD.73.124011}{124011}

\bibitem{Boyd_1987}
Boyd J~P 1987 {Spectral methods using rational basis functions on an infinite
  interval} \emph{Journal of Computational Physics} \textbf{69} 112--142

\bibitem{Boyd_2000}
Boyd J~P 2001 \emph{{Chebyshev and Fourier Spectral Methods}} (Dover
  Publications Inc., New York)

\bibitem{lapack_users_guide}
Anderson E, Bai Z, Bischof C, Blackford S, Demmel J, Dongarra J, Du~Croz J,
  Greenbaum A, Hammarling S, McKenney A and Sorensen D 1999 \emph{{LAPACK}
  Users' Guide} (Philadelphia, PA: Society for Industrial and Applied
  Mathematics) third edn. ISBN 0-89871-447-8 (paperback)

\bibitem{einsteintoolkit_web}
{Einstein Toolkit}: Open software for relativistic astrophysics
  (\url{http://einsteintoolkit.org/})

\bibitem{Loffler_2011}
L{\"{o}}ffler F, Faber J, Bentivegna E, Bode T, Diener P, Haas R, Hinder I,
  Mundim B~C, Ott C~D, Schnetter E, Allen G, Campanelli M and Laguna P 2012
  {{T}he {E}instein {T}oolkit: {A} {C}ommunity {C}omputational {I}nfrastructure
  for {R}elativistic {A}strophysics} \emph{Class. Quantum Grav.}
  \textbf{29}(11) 
  \href{http://iopscience.iop.org/article/10.1088/0264-9381/29/11/115001}{115001}

\bibitem{cactuscode_web}
{Cactus Computational Toolkit} (\url{http://www.cactuscode.org/)}

\bibitem{goodale_2002}
Goodale T, Allen G, Lanfermann G, Mass{\'o} J, Radke T, Seidel E and Shalf J
  2003 The {Cactus} framework and toolkit: Design and applications \emph{Vector
  and Parallel Processing -- VECPAR'2002, 5th International Conference, Lecture
  Notes in Computer Science} (Berlin: Springer) (\url{http://edoc.mpg.de/3341})

\bibitem{schnetter_2003}
Schnetter E, Hawley S~H and Hawke I 2004 {Evolutions in 3-D numerical
  relativity using fixed mesh refinement} \emph{Class. Quantum Grav.}
  \textbf{21} 
  \href{https://iopscience.iop.org/article/10.1088/0264-9381/21/6/014}{1465--1488}

\bibitem{berger_1984}
Berger M~J and Oliger J 1984 {Adaptive Mesh Refinement for Hyperbolic Partial
  Differential Equations} \emph{J. Comput. Phys.} \textbf{53} 484

\bibitem{mclachlan_web}
{McLachlan}, a public {BSSN} code
  (\url{http://www.cct.lsu.edu/~eschnett/McLachlan/})

\bibitem{kranc_web}
{Kranc}: {Kranc} assembles numerical code (\url{http://kranccode.org/})

\bibitem{brown_2008}
Brown J~D, Diener P, Sarbach O, Schnetter E and Tiglio M 2009 {Turduckening
  black holes: an analytical and computational study} \emph{Phys. Rev. D}
  \textbf{79} 
  \href{https://journals.aps.org/prd/abstract/10.1103/PhysRevD.79.044023}{044023}

\bibitem{alcubierre_1999}
Alcubierre M, Brandt S, Br\"ugmann B, Holz D, Seidel E, Takahashi R and
  Thornburg J 2001 {Symmetry without symmetry: Numerical simulation of
  axisymmetric systems using Cartesian grids} \emph{Int. J. Mod. Phys.}
  \textbf{D10} 
  \href{https://www.worldscientific.com/doi/abs/10.1142/S0218271801000834}{273--290}

\bibitem{bicgstab}
van~der Vorst H~A 1992 Bi-cgstab: A fast and smoothly converging variant of
  bi-cg for the solution of nonsymmetric linear systems \emph{SIAM Journal on
  Scientific and Statistical Computing} \textbf{13}(2) 
  \href{http://dx.doi.org/10.1137/0913035}{631--644}

\bibitem{szilagyi_2009}
Szil\'agyi B, Lindblom L and Scheel M~A 2009 Simulations of binary black hole
  mergers using spectral methods \emph{Phys. Rev. D} \textbf{80} 
  \href{https://link.aps.org/doi/10.1103/PhysRevD.80.124010}{124010}

\end{thebibliography}

\end{document}